\documentclass[12pt,a4paper]{article}
\usepackage[utf8]{inputenc}
\usepackage[english]{babel}
\usepackage{amsmath, amsfonts, amssymb}
\usepackage{graphicx}
\usepackage{booktabs}
\usepackage{array}
\usepackage{xcolor}
\usepackage{colortbl}
\usepackage{multirow}
\usepackage{url}
\usepackage[none]{hyphenat} 
\usepackage{hyperref}
\usepackage{cite}
\usepackage{bm}
\usepackage{geometry}
\usepackage{fancyhdr}
\usepackage{graphicx}
\usepackage{float}      
\usepackage{placeins}
\usepackage{geometry}
\usepackage{setspace}
\usepackage{times} 
\usepackage[font=small,labelfont=bf] {caption} 
\usepackage{hyperref}
\hypersetup{
    colorlinks=true,
    linkcolor=blue,     
    citecolor=blue,     
    filecolor=blue,     
    urlcolor=blue       
}

\begin{document}
{\centering
\fontsize{16pt}{20pt}\selectfont \bfseries
On-chip microwave sensing of quasiparticles \\in
$\alpha$-tantalum superconducting circuits on silicon \\for scalable quantum technologies \par
}

\vspace{1em}
{\centering
\fontsize{12pt}{14pt}\selectfont
Shima Poorgholam-Khanjari, Paniz Foshat,\\
Mingqi Zhang, Valentino Seferai, Martin Weides, and Kaveh Delfanazari$^*$ \par
}

\vspace{1em}

{\centering
\fontsize{11pt}{13pt}\selectfont
\textit{Electronics and Nanoscale Engineering Division, James Watt School of Engineering,\\
University of Glasgow, Glasgow, UK} \par
}

\vspace{0.5em}
{\centering
\fontsize{10pt}{12pt}\selectfont
$^*$Corresponding author: \href{mailto:kaveh.delfanazari@glasgow.ac.uk}{kaveh.delfanazari@glasgow.ac.uk} \par
}

\vspace{1.5em}

\renewenvironment{abstract}{
    \noindent\textbf{Abstract:} \\
    
}{\par\vspace{1em}}

\section*{Abstract}
The performance and scalability of superconducting quantum circuits are fundamentally constrained by non-equilibrium quasiparticles, which induce microwave losses that limit resonator quality factors and qubit coherence times. Understanding and mitigating these excitations is therefore central to advancing scalable quantum technologies. Here, we demonstrate on-chip microwave sensing of quasiparticles in high-$Q$ $\alpha$-tantalum coplanar waveguide resonators on silicon, operated in the single-photon regime. Temperature-dependent measurements reveal persistent non-equilibrium quasiparticles at millikelvin temperatures, producing a measurable suppression of the internal quality factor ($Q_i$) relative to theoretical expectations. By benchmarking across materials, we find that the quasiparticle density in $\alpha$-Ta is approximately one-third that of NbN at equivalent normalised temperatures ($T/T_c$), directly correlating with reduced microwave loss. Our methodology establishes a scalable platform for probing quasiparticle dynamics and points towards new routes for engineering superconducting circuits with improved coherence, with impact on qubit readout resonators, kinetic-inductance detectors, and emerging quantum processors and sensors.\\

\textit{\textbf{Key words:}} Tantalum, superconducting microwave coplanar waveguide resonators, quasiparticle, internal quality factor, qubit, coherence, scalable quantum computing.

\section{Introduction}
In recent decades, superconducting qubits have been one of the most intriguing research subjects due to their potential applications in quantum information processing \cite{clarke2008superconducting,wei2020compact,kroll2019magnetic}.
A significant portion of the area in superconducting quantum circuits is typically occupied by superconducting resonators \cite{carter2019low,goetz2016loss,goppl2008coplanar,ohya2013sputtered,foshat2025characterizing}. 
In fact, high-$Q$ \cite{poorgholam2025engineering,zikiy2023high,lozano2022manufacturing,blair1987high} superconducting coplanar waveguide (CPW) resonators with low microwave loss are essential components of quantum computation \cite{wu2021strong,jurcevic2021demonstration}. 
In addition to superconducting resonators and qubits, recent advancements in hybrid superconductor–semiconductor circuits have opened new opportunities for scalable quantum technologies \cite{delfanazari2018chip,delfanazari2024quantized,pitsun2020cross}. These systems exploit the interplay between superconductivity and tunable electronic structures to generate novel quantum transport phenomena \cite{delfanazari2024large,delfanazari2017chip,serra2020evidence}. These advancements highlight the growing demand for materials that demonstrate robust superconducting characteristics and minimum energy dissipation.

Superconducting devices are strongly affected by excess quasiparticles at low temperatures, particularly in applications such as quantum processors and superconducting resonators, where quasiparticle-induced losses degrade performance. Understanding and mitigating non-equilibrium quasiparticles is therefore essential to advancing superconducting technologies \cite{de2011number,serniak2018hot,gao2008equivalence,fischer2023nonequilibrium}. When a photon with energy significantly well above $2\Delta$ interacts with a superconductor, it can break a Cooper pair into two high-energy quasiparticles with opposite spins. 
These quasiparticles then decay by emitting phonons, which can break additional pairs into even more quasiparticles with lower energy. 
This process generates a large number of quasiparticles \cite{catelani2022using,de2011number}. 
The electromagnetic response of superconductors is also significantly influenced by quasiparticles. 
Moreover, the performance of a variety of superconducting circuits is degraded by non-equilibrium quasiparticle excitations \cite{serniak2018hot,marin2020active}. 
At very low temperatures ($T \ll T_c$), the number of thermally excited quasiparticles should be extremely small. 
However, recent measurements have shown that the quasiparticle density at low temperatures exceeds the expected thermal equilibrium value by orders of magnitude \cite{de2011number,vool2014non,bespalov2016theoretical}. 
This excess of quasiparticles, called quasiparticle poisoning \cite{aumentado2004nonequilibrium,bespalov2016theoretical,anthony2024stress}, affects the performance of superconducting devices. Tantalum (Ta) has become a predominant material for superconducting circuitry, as it provides a relatively high superconducting transition temperature and minimal intrinsic dissipation, making it a reliable choice for quantum devices \cite{lozano2024low,lozano2025reversing,marcaud2025low,crowley2023disentangling,urade2024microwave,jia2023investigation}.

In this work, we investigate the effect of non-equilibrium quasiparticles in superconducting tantalum ($\alpha$-Ta) CPW resonators in the single-photon regime, over a temperature range of 0.77-1 K, and compare the results with our recent work on NbN-based circuits \cite{foshat2025characterizing}.
Furthermore, we experimentally determine and theoretically model the quasiparticle density ($n_{qp}$), showing that $n_{qp}$ persists at low temperatures ($T \ll T_c$). In order to characterise the crystalline phase, grain structure, and film quality that are the foundation of the device performance, we use transmission electron microscopy (TEM) and X-ray diffraction (XRD). 
Furthermore, we use a conventional approach to model TLS loss \cite{pappas2011two,crowley2023disentangling} and compute the complex conductivity of the Ta film using the Mattis--Bardeen theory \cite{mattis1958theory} to quantify the contribution of quasiparticle dissipation.

\section { Microscopy and Deep Cryogenic Microwave Spectroscopy}

\begin{figure}[!ht]
  \centering
\begin{tabular}{c c}
\includegraphics[width=0.3\textwidth,height=4.5cm]{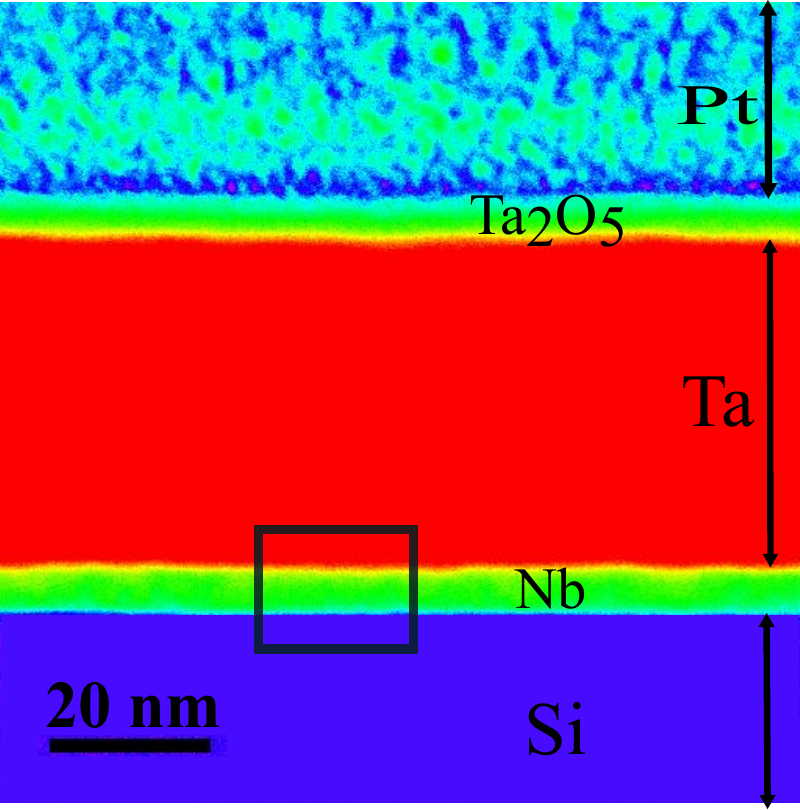} &   \includegraphics[width=0.3\textwidth,height=4.5cm]{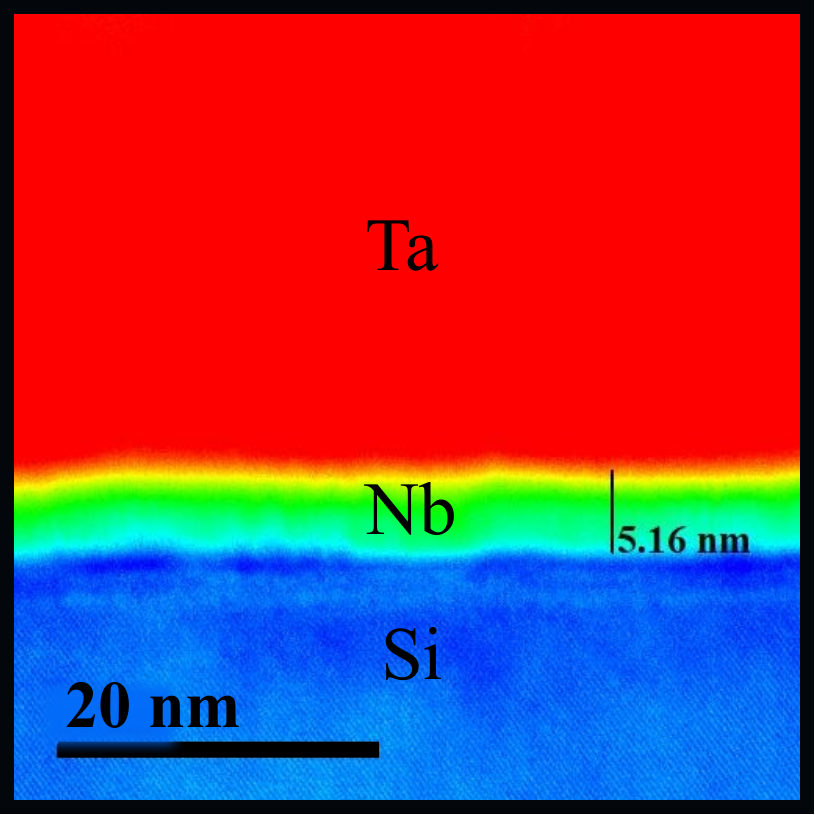}\\
(a) & (b)\\
      \end{tabular}
\begin{tabular}{c c}
\includegraphics[width=0.3\textwidth,height=4.5cm]{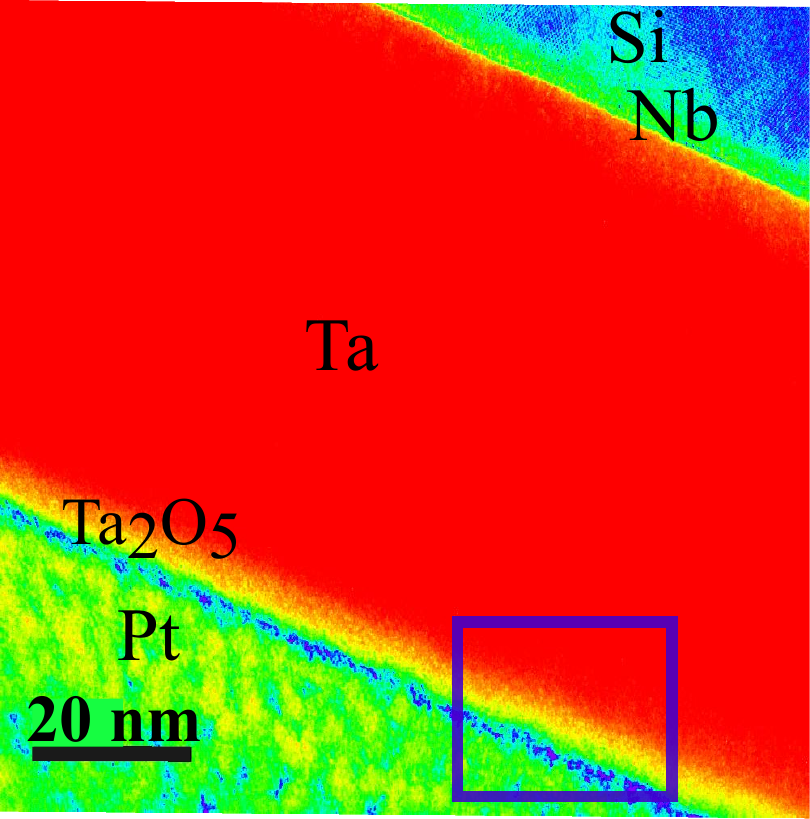}&
\includegraphics[width=0.3\textwidth,height=4.5cm]{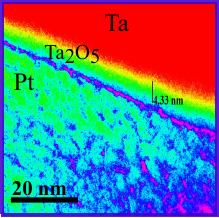}\\
(c) &(d) 
 \end{tabular}
\begin{figure}[H]
   \centering
\includegraphics[width=0.6\textwidth,height=6cm]{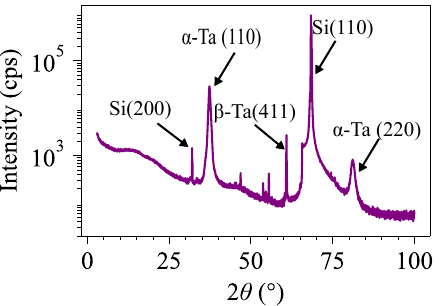}\\
\hspace{24pt}
(e)
    \caption{(a) Cross-sectional dark-field STEM image of a bare Ta film with an Nb seed layer on a Si substrate. (b) Zoomed-in view of STEM image of the area selected in (a), which shows the thickness of the Nb layer. (c) TEM image of the sample. (d) Zoomed-in view of TEM image of the area selected in (c) showing the thickness of Ta oxide. (e) XRD diffraction pattern of 40 nm $\alpha-$Ta film on Si substrate with an Nb seed layer.   }
    \label{STEM}
\end{figure}
\end{figure}

The fabrication process was initiated with wafer cleaning, followed by deposition of a tantalum (Ta) layer of different thicknesses (here we mostly focus on a 40 nm Ta film) with a niobium (Nb) seed layer on a high-resistivity silicon (Si) substrate using DC sputtering at room temperature. Prior to device patterning, a bare 40 nm  $\alpha$-tantalum film was selected and its microstructure and structural integrity were characterised. TEM was employed to investigate the crystalline structure, grain size, and interface quality between the Ta layer, the high-resistivity Si
substrate and the Nb seed layer. Figure \ref{STEM}(a) presents a high-resolution cross-sectional scanning transmission electron microscope (STEM) image of a bare Ta film that was deposited on a Si substrate, with an Nb seed layer. It highlights the structural quality of the multilayer stack as well as the interface sharpness. To protect the underlying film from milling damage, a focused ion beam (FIB) protective coating was applied during TEM sample preparation and shown in the dark-field STEM image as the top platinum (Pt) layer (Fig. \ref{STEM}(a)). 

The Ta film exhibits a thin native tantalum oxide ($\text{Ta}_2\text{O}_5$) layer on the surface, followed by the Ta layer, an ultrathin Nb seed layer, and a crystalline Si substrate. Figure \ref{STEM}(b) presents a magnified STEM image that verifies the deposition process by precisely measuring the Nb layer thickness at 5.16 nm. Figure \ref{STEM}(c) and (d) show bright-field TEM images and an enlarged view of the Pt/$\text{Ta}_2\text{O}_5$/Ta interface, respectively. These images reveal that the layers are sharply separated at the atomic level, with almost no interdiffusion or defect formation. Moreover, the thickness of the tantalum oxide layer ($\text{Ta}_2\text{O}_5$) is about 4.5 nm, which is shown in Fig. \ref{STEM}(d). This thin oxide layer is consistent with minimal losses, which supports the high internal quality factor of Ta films, and demonstrates the feasibility of fabricating Ta/Nb/Si heterostructures with nanometer-scale control over thickness, composition, and interface quality, which is essential for the development of robust quantum devices and superconducting circuits.

\begin{figure}[H]
    \centering
    \begin{tabular}{cc}
        \includegraphics[width=0.48\linewidth, height=5.5cm]{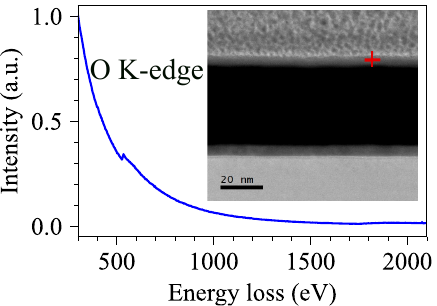}&
    \includegraphics[width=0.48\linewidth, height=5.5cm]{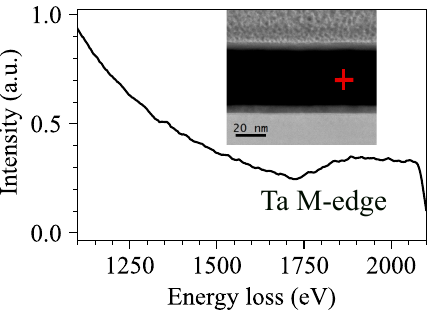} 
         \\
        (a) & (b)   
    \end{tabular}
    
    \begin{tabular}{cc}
    \includegraphics[width=0.48\linewidth, height=5.5cm]{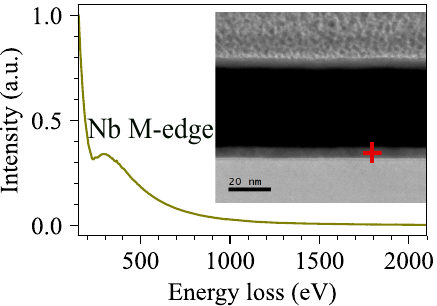}
 & 
\includegraphics[width=0.48\linewidth, height=5.5cm]{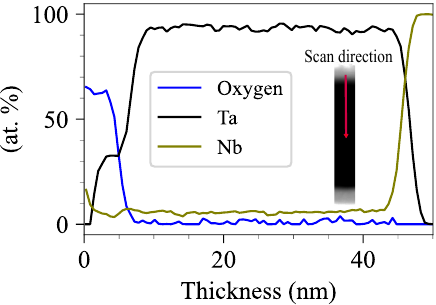} \\
(c) & (d)
        \end{tabular}
    \caption{STEM–EELS analysis of a 40 nm Ta device. (a) O K-edge, (b) Ta M-edge, and (c) Nb M-edge spectra collected from the indicated regions (insets). (d) Corresponding elemental (atomic) concentration profile (O, Ta, Nb) extracted along the scan direction, confirming the spatial distribution of the oxide, Ta, and Nb layers. The red cross indicates the beam position during scanning.}
    \label{EELS}
\end{figure}
In order to confirm the film’s crystalline phase composition, X-ray diffraction (XRD) measurements were performed as part of the structural investigation. While the TEM imaging directly probed the particle morphology and interface quality, XRD analysis provided complementary information on the phase identity and crystallographic orientation. The XRD pattern of the 40~nm Ta film deposited on Si is shown in Fig. \ref{STEM}(e). Two distinct peaks at approximately $38^\circ$ and $70^\circ$ correspond to $\alpha$-Ta~(110) and $\alpha$-Ta~(220) reflections, respectively, confirming the presence of the stable body-centered cubic (bcc) phase. In addition, a peak at $\sim 60^\circ$ is attributed to the $\beta$-Ta~(411) reflection, indicating the coexistence of the metastable tetragonal $\beta$ phase, which is commonly observed in sputtered Ta films. In addition, the peaks arising from the silicon substrate are also observed at $\sim 28^\circ$ [Si~(200)] and $\sim 69^\circ$ [Si~(110)]. The presence of both $\alpha$-Ta and $\beta$-Ta reflections suggests partial phase transformation during deposition, with $\alpha$-Ta being the dominant phase. The lack of additional tantalum oxide-related peaks suggest that any oxide layer is either amorphous or below the XRD detection limit.
Figure \ref{EELS} shows the chemical and structural analysis of the Ta/Nb/Si stack. Figs. \ref{EELS}(a)–(c) illustrate the electron energy loss spectroscopy (EELS) spectra that were obtained at representative regions of the film, which correspond to the O K-edge, Ta M-edge, and Nb M-edge, respectively. The different starting points and relative strengths of these edges show that there is oxygen on the surface of the film, metallic Ta throughout the bulk of the deposited layer, and Nb only in the interfacial area. Fig. \ref{EELS}(d) shows the measured compositional depth profile across the multilayer. A thin region at the top surface that is rich in oxygen (about 5 nm) is seen, which is in line with the creation of a native $\text{Ta}_2\text{O}_5$ layer. Under this oxide, the profile is mostly made up of metallic Ta, which is around 40 nm thick. A $\sim$ 5 nm Nb layer can be observed at the interface with the substrate before the signal drops into the underlying Si (not shown). The sharp transitions in elemental composition across the interfaces highlight the structural integrity of the stack and confirm that interdiffusion is minimal. Therefore, the expected Si/ Nb/Ta/$\text{Ta}_2\text{O}_5$ structure was fabricated and preserved during subsequent analysis.

\begin{figure}[H]
  \centering
\begin{tabular}{c c}
\includegraphics[width=0.45\textwidth,height=5cm]{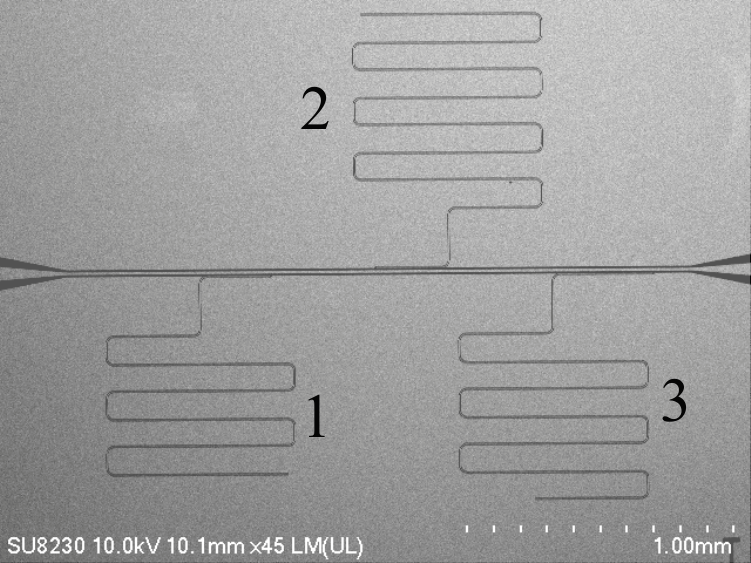} &  

\includegraphics[width=0.45\textwidth,height=5cm]{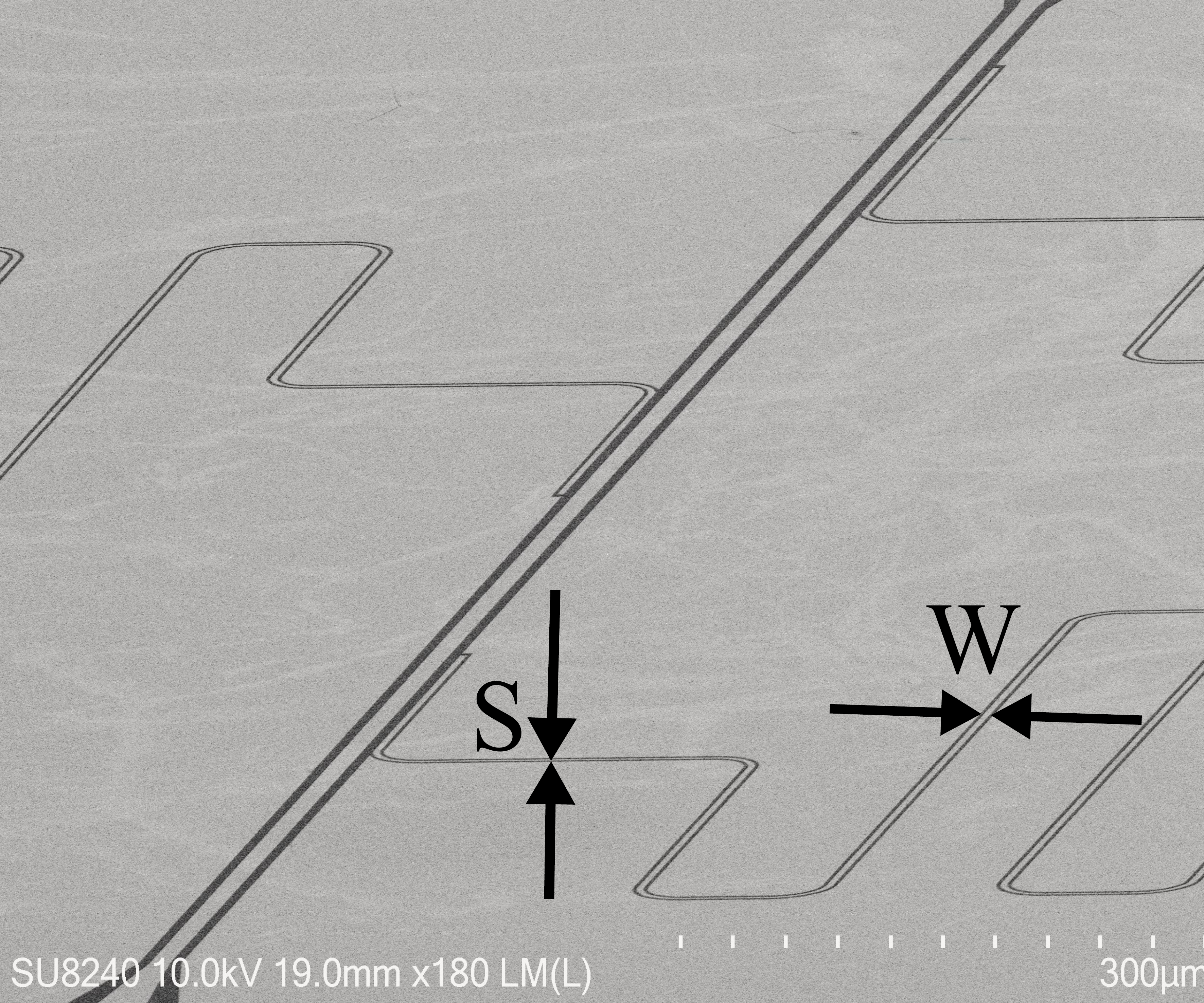} \\
(a) & (b)

     \end{tabular}
     \includegraphics[width=0.94\textwidth,height=4cm]{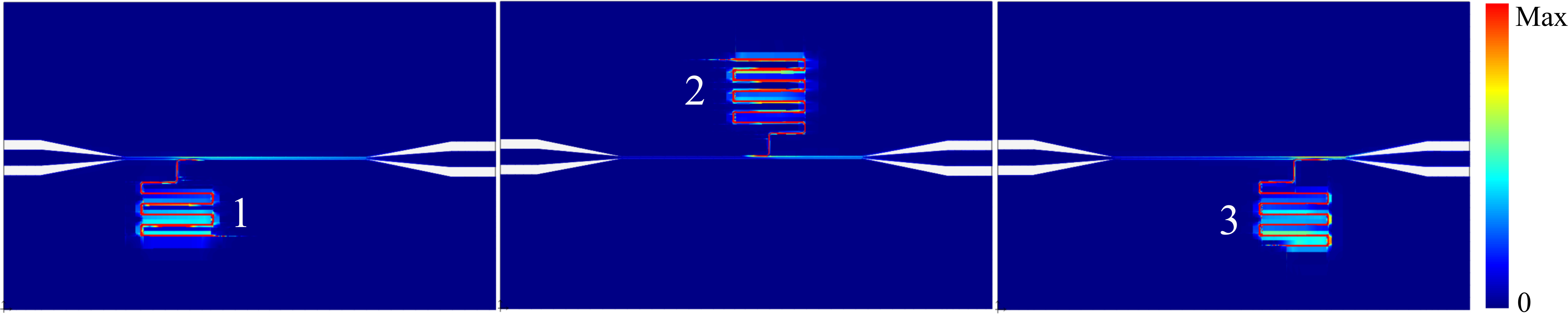}\\
     \centering
     (c)
    \caption{(a) Top view SEM image of superconducting Ta chip with three CPW resonators coupled to a transmission line. (b) Zoomed-in view of SEM image of the Ta circuit with W=4 $\mu m$ and S=2 $\mu m$. (c) The surface current density magnitude $|\mathbf{J}_s|$ (A/m) for three typical resonators at their resonance frequencies.}
    \label{SEM}
\end{figure}


The device follows the design and fabrication methods reported in \cite{poorgholam2025engineering}. The device consists of three quarter-wavelength resonators coupled to a common transmission line. 
E-beam lithography was used for patterning, followed by dry etching of the sample with CF$_4$/Ar gases. The scanning electron micrograph (SEM) image of the sample is shown in Fig. \ref{SEM}(a) and (b).
Afterwards, the sample was diced into $5 \times 5 \ \mathrm{mm}^2$ chips. 
One chip was chosen, wire-bonded to a copper sample box, and mounted in an Oxford Instruments Triton 200 Dilution Refrigerator (DR) system, cooled to a base temperature of $T=77$ mK.

\begin{figure}[H]
\begin{tabular}{cc}
    \centering
    
\includegraphics[width=0.48\linewidth, height=6.1cm]{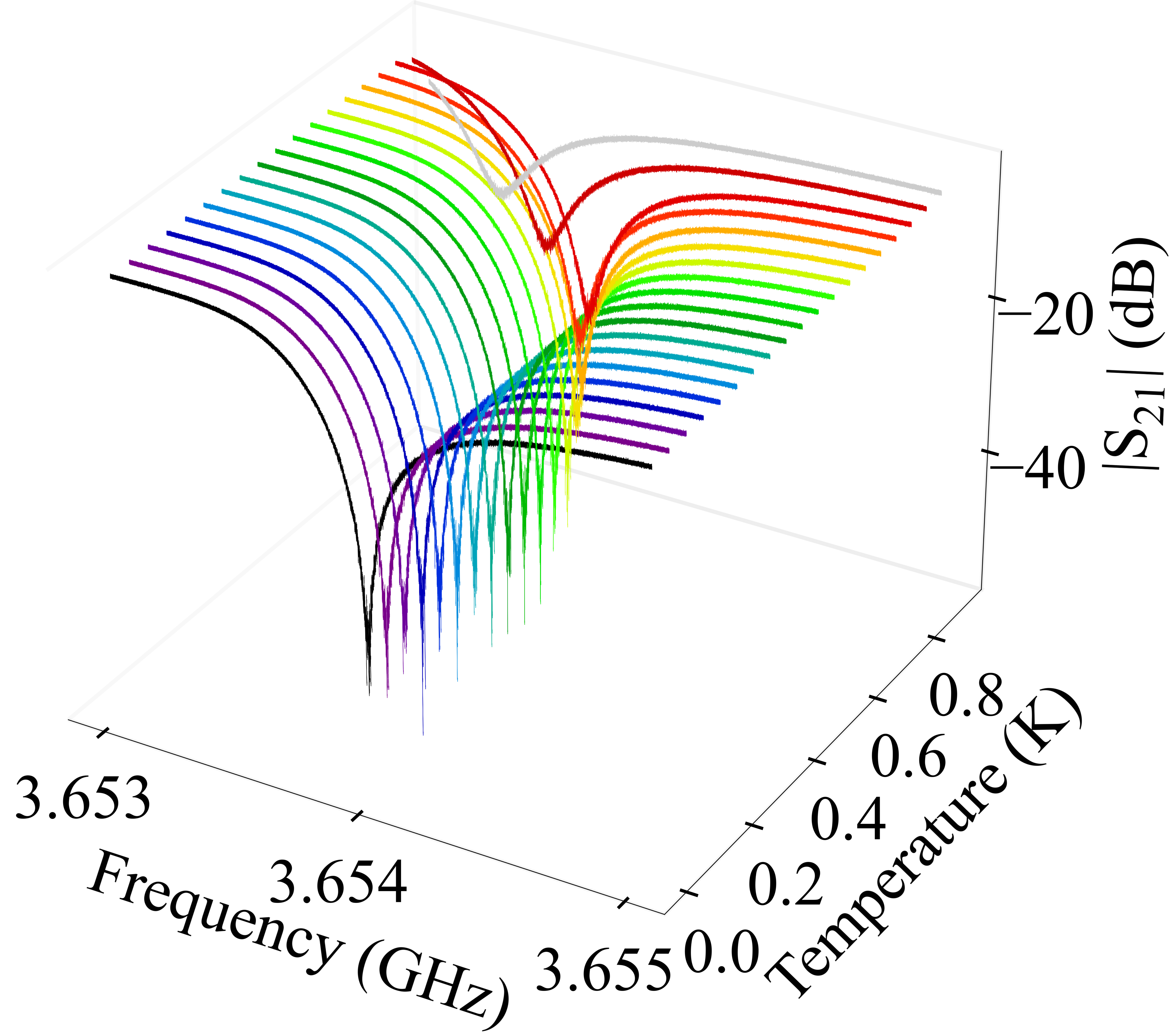}&

\includegraphics[width=0.48\linewidth, height=6.1cm]{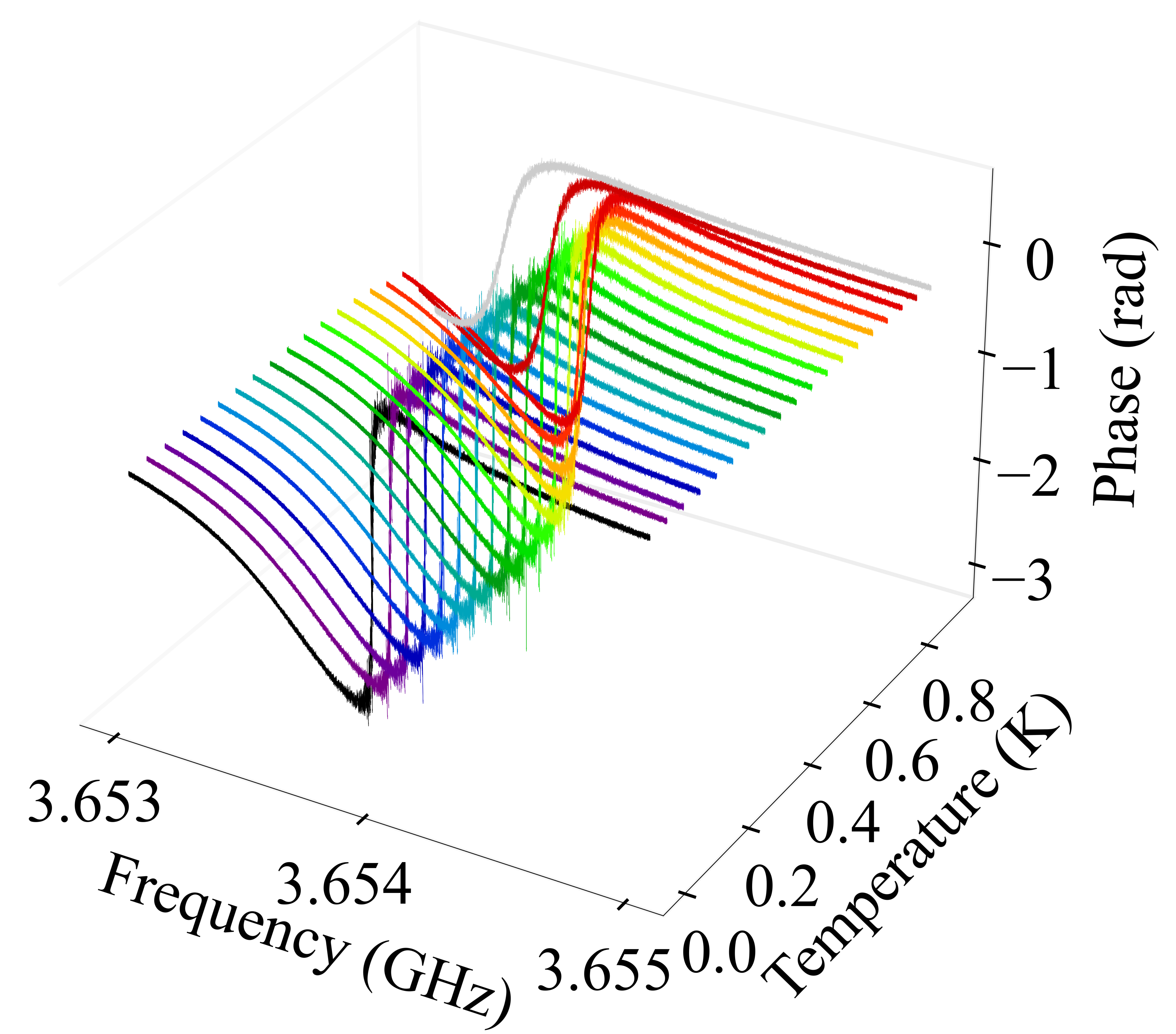}\\
(a) & (b)
    \end{tabular}

\begin{tabular}{cc}

\includegraphics[width=0.48\linewidth, height=5.5cm]{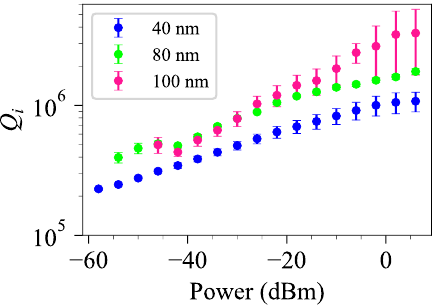}&

\includegraphics[width=0.48\textwidth,height=5.5cm]{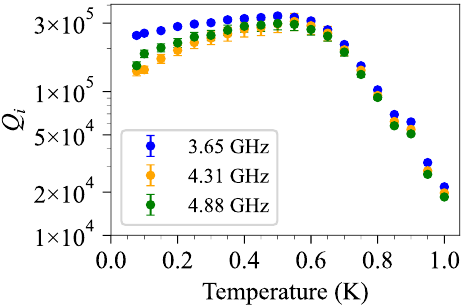}\\
(c) & (d)
 \end{tabular}
 \begin{tabular}{c c}
      \includegraphics[width=0.48\textwidth,height=5.5cm]{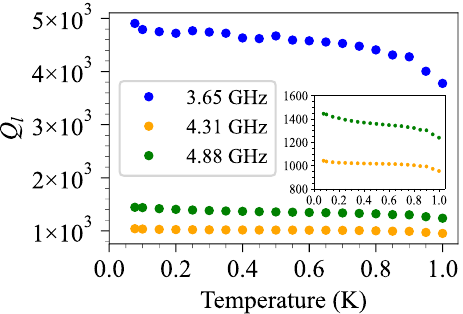}&
      \includegraphics[width=0.48\linewidth, height=5.5cm]{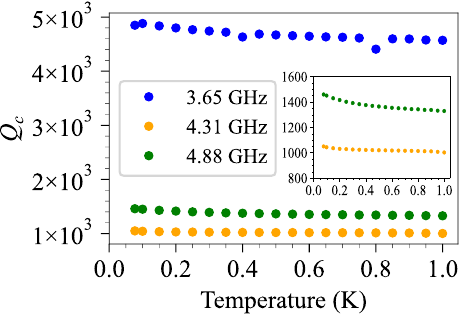}\\
      (e) & (f)
 \hspace{1.4cm} 
 \end{tabular} 
    \caption{3-D view of measured amplitude (a), and phase (b) of the resonator at $f_r$=3.65 GHz at different temperatures from $T$ = 77 mk to $T$ = 1 K. (c) Internal quality factor ($Q_i$) of three (different thicknesses) Ta CPW resonators on Si as a function of power at $T$=77 mK. (d) Internal quality factor ($Q_i$) of the 40 nm CPW resonator on Si as a function of temperature at single photon regime. (e) Loaded quality factor ($Q_l$) vs temperature, and (f) Coupling quality factor ($Q_c$) vs temperature for the 40 nm Ta CPW resonator. Insets are the zoomed-in areas for presentation purposes.}
    \label{temperaturesweep}
\end{figure}

The input signals from the vector network analyser (VNA) were attenuated by 20 dB at room temperature and by an additional 60 dB inside the refrigerator before reaching the transmission line of the superconducting circuit. The output signals from the device were first amplified by a 40 dB low-noise high-electron mobility transistor (HEMT) amplifier at the 4 K stage, and subsequently by a room-temperature amplifier with 45 dB gain. Figure \ref{temperaturesweep}(a) shows a 3-D view of the measured amplitude as a function of frequency in the single photon regime at temperatures between 77 mK and 1K, at $f_r$= 3.654 GHz. It can be seen that the resonance frequency shifts to lower frequencies as the temperature increases from 550 mK to 1K.  Figure \ref{temperaturesweep}(b) shows a 3-D view of the measured phase as a function of frequency in the single photon regime, which repeats the behaviour of amplitude. Figure \ref{temperaturesweep}(c) shows the power dependence of $Q_i$ for three different Ta samples with thicknesses of 40 nm, 80 nm, and 100 nm, with the highest $Q_i$ of $3\times10^{6}$ for the 100 nm Ta film. Figure \ref{temperaturesweep} (d) demonstrates the temperature dependence of $Q_i$ for all resonance frequencies of a Ta= 40 nm CPW resonator, while Fig. \ref{temperaturesweep}(e) and (f) show the $Q_l$ and $Q_c$, respectively. For $T \lesssim 0.5~\mathrm{K}$, $Q_l$ remains nearly constant and is consistent with $Q_c$, suggesting that the system is operating in the coupling-limited regime. In this regime, dissipation is primarily influenced by external coupling to the feedline, and intrinsic losses are negligible. Above $\sim0.5~\mathrm{K}$, $Q_l$ decreases below $Q_c$, signifying a transition to the loss-limited regime dominated by thermally generated quasiparticles. Moreover, the weak temperature dependence of $Q_l$, despite the dramatic $Q_i$ variations, validates the overcoupled design approach for achieving temperature-stable superconducting resonator performance.


\subsection{Modelling Complex Conductivity and Quasiparticle Dynamics}
The concept of complex conductivity $\sigma(T) = \sigma_1(T) - j\sigma_2(T)$ was first introduced by Glover and Tinkham \cite{glover1957conductivity} for the superconducting states. For the calculation of quasiparticle density $n_{qp}(T)$, first we need to obtain complex conductivity. By considering the condition of $\hbar\omega \ll \Delta_0$ and $k_B T \ll \Delta_0$, the Mattis–Bardeen relations are expressed as \cite{gao2008equivalence,gao2008physics}: 
\begin{equation}
\frac{\sigma_1(T)}{\sigma_n} = \frac{4\Delta_0}{\hbar\omega} e^{-\frac{\Delta_0}{k_B T}} \sinh\left(\frac{\hbar\omega}{2k_B T}\right) K_0\left(\frac{\hbar\omega}{2k_B T}\right)
\label{sigma1/sigman}
\end{equation}

\begin{equation}
\frac{\sigma_2(T)}{\sigma_n} = \frac{\pi\Delta_0}{\hbar\omega} \left[1 - \sqrt{\frac{2\pi k_B T}{\Delta_0}} e^{-\frac{\Delta_0}{k_B T}} - 2e^{-\frac{\Delta_0}{k_B T}} e^{-\dfrac{\hbar\omega}{2k_B T}} I_0\left(\dfrac{\hbar\omega}{2k_B T}\right)\right]
\label{sigma2/sigman}
\end{equation}

\begin{equation}
n_{qp}(T) \approx 2N_0\sqrt{2\pi k_B T \Delta(T)} e^{-\frac{\Delta(T)}{k_B T}}
\end{equation}

\begin{equation}
\Delta_0=1.76\times K_B \times T_c  
\end{equation}


Where $\sigma_1$ is the real part of conductivity, $\sigma_2$ is the imaginary part of conductivity, 
$\sigma_n$ is the normal-state conductivity, $\Delta_0$ is the superconducting energy gap at $T=0~\mathrm{K}$, 
$\hbar$ is the reduced Planck’s constant, $k_B$ is Boltzmann’s constant, 
$N_0$ is the density of states at the Fermi level, which for Ta is 
$N_0 \approx 6.9 \times 10^{28}~\mathrm{states}/(\mathrm{m}^3~\mathrm{eV})$ \cite{guruswamy2018nonequilibrium,claeson1974microscopic}, 
$T_c \approx 4.06~\mathrm{K}$ for 40 nm Ta, $I_0$ and $K_0$ are the modified Bessel functions of the first 
and the second kind, respectively.

\begin{figure}
    \centering
\begin{tabular}{cc}
\includegraphics[width=0.48\textwidth,height=5.5cm]{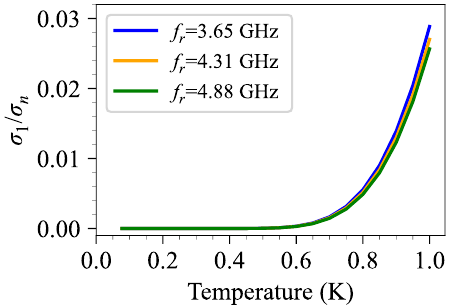} &
\includegraphics[width=0.48\textwidth,height=5.5cm]{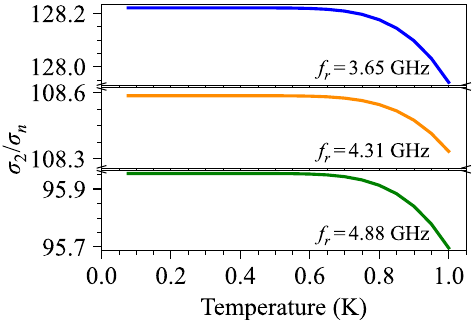} \\
     (a) & (b)
    \end{tabular}
    \caption{(a) The calculated real part of the complex conductivity, $\sigma_1$, as a function of temperature. (b) The calculated imaginary part of the complex conductivity, $\sigma_2$, as a function of temperature for the Ta superconducting CPW resonator. Both plots are calculated from all the measured resonance frequencies.}
    \label{conductivity}
\end{figure}

Using the above equations, the complex conductivity was obtained, where the real part 
$\sigma_1(T)$ represents losses caused by quasiparticles, resulting in energy dissipation in the resonator, 
and the imaginary part $\sigma_2(T)$ represents the inductive response of superconducting Cooper pairs. The latter is the determining factor in the superconductor's ability to store and transfer energy without dissipation and is directly related to the kinetic inductance $L_k$ of the superconductor. 
Figure \ref{conductivity}(a) and (b) show the calculated real and imaginary parts of the complex conductivity as a function of temperature.  At low temperatures, the thermal energy is insufficient to beak a significant number of Cooper pairs, 
leading to a limited quantity of quasiparticles $n_{qp}(T)$. 
In fact, $\sigma_1(T) \propto n_{qp}(T) \propto e^{\frac{-\Delta}{k_B T}}$ which is extremely small at low temperatures as $T \to 0$, where $\Delta(T) \to \Delta_0$. This is due to the fact that $\Delta_0$ is significantly larger than $k_B T$, resulting in the presence of only a small number of quasiparticles. 
Following this, as the temperature increases, the conductivity increases, which in turn increases the density of quasiparticles. However, Fig. \ref{conductivity}(b) illustrates a downward trend. 
To understand this, consider the relationship 
$ L_k \propto \frac{1}{\sigma_2(T)}.$
As the temperature rises, the density of quasiparticles increases, leading to an increase in kinetic inductance and a corresponding reduction in $\sigma_2(T)$. 
As a result, the contribution of quasiparticle loss in superconducting CPW resonators can be determined by calculating the complex conductivity of the superconductor film.
\subsection{Thermal and Non-equilibrium Quasiparticle Density}
\begin{figure}[!htb]
    \centering
\begin{tabular}{cc}
\includegraphics[width=0.48\textwidth,height=5cm]{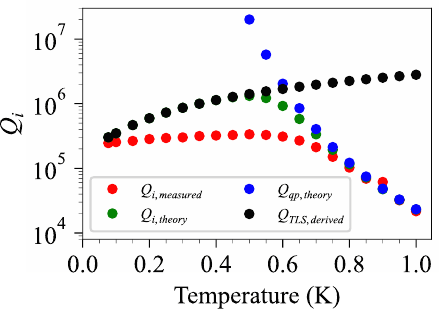} &
\includegraphics[width=0.48\textwidth,height=5.1cm]{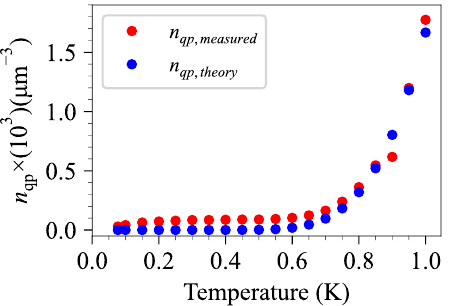} \\
     (a) & (b)
    \end{tabular}
\begin{tabular}{cc}
\includegraphics[width=0.48\textwidth,height=5.5cm]{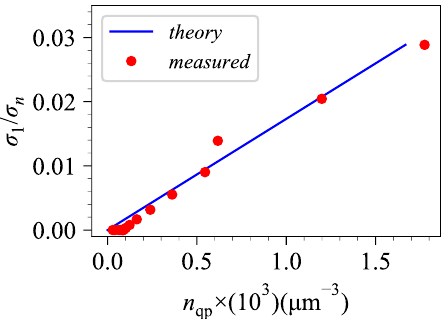} &
\includegraphics[width=0.48\textwidth,height=5.5cm]{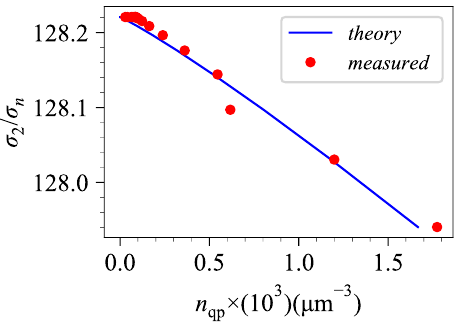}\\ 
     (c) & (d)
     \end{tabular}
    \caption{(a) $Q_{i,measured}$  and $Q_{i,theory}$ versus temperature at single photon regime ($\langle n_{ph} \rangle \sim 1$) with the theoretical model of TLS and quasiparticle loss. Both plots are calculated and measured at $f_r$ = 3.65 GHz. (b) Theoretical and measured quasiparticle density of Ta CPW resonator. 
(c) Normalised real part of the conductivity, $\sigma_{1}/\sigma_{n}$, versus $n_{\mathrm{qp}}$ showing both theoretical prediction (blue line) and experimental data (red circles). 
(d) Normalised imaginary part of the conductivity, $\sigma_{2}/\sigma_{n}$, versus $n_{\mathrm{qp}}$ with corresponding theory and  (all data are for 40 nm thickness Ta).} 

    \label{Qi-measured-theory}
\end{figure}

This section presents the observed and theoretical values of quasiparticle density, with a discussion of 
$Q_{i,{theory}}$, $Q_{i,measured}$, $Q_{qp,theory}$, and $Q_{TLS,derived}$. 
The loss model for quasiparticles is defined as  \cite{poorgholam2025engineering}:
\begin{equation}
\delta_{qp}(T) = \frac{1}{Q_{qp}} 
= \frac{\alpha}{\pi} \sqrt{\frac{2\Delta(T)}{h f_r}} 
\cdot \frac{n_{qp}(T)}{N_0 \Delta(T)}.
\end{equation}
Where $\alpha$ is the ratio of kinetic inductance to total inductance.
\begin{equation}
\delta_i = \delta_{TLS}(T, P) + \delta_{qp}(T) + \delta_{other}
\end{equation}

\begin{equation}
\delta_{qp,measured}(T) = \frac{1}{Q_{i,measured}} - \frac{1}{Q_{TLS,derived}}
\label{delta-qp-measured}
\end{equation}
We can rewrite the Eq.\ref{delta-qp-measured} to obtain $n_{qp,theory}$:
\begin{equation}
n_{qp,measured}(T) \approx \delta_{qp,measured}(T) N_0 \Delta(T) \frac{\pi}{\alpha} \sqrt{\frac{h f_r}{2\Delta(T)}}
\label{n-qp-measured}
\end{equation}
By plotting $Q_i$ versus the photon number $\langle n_{ph} \rangle$ and fitting using: 
\begin{equation}
    \delta_{TLS}(T,P) = \frac{1}{Q_{TLS}} 
= \frac{1}{Q_{TLS}^0} 
\tanh \left( \frac{h f_r}{2 k_B T} \right) 
\bigg/ \sqrt{1 + \left( \frac{n_{ph}}{n_c} \right)^{\beta}},
\end{equation}

the values of 1/${Q_{TLS}^0}$, $n_c$, and $\beta$ are obtained \cite{poorgholam2025engineering}, 
which facilitates the derivation of $\delta_{TLS}(T,P)$. 
Then, by substituting Eqs. \eqref{delta-qp-measured} and \eqref{n-qp-measured},
$\delta_{qp,measured}(T)$ and $n_{qp,measured}(T)$ are obtained which are shown in Fig. \ref{Qi-measured-theory}(a) and (b). 

The surface impedance $Z_s$ of a superconductor can be obtained  \cite{gao2008physics}:
\begin{equation}
Z_s = \sqrt{\frac{j\mu_0 \omega}{\sigma_1(T) - j\sigma_2(T)}} = R_s + j\omega L_s
\label{surface impedance}
\end{equation}
By substituting Eqs.  \eqref{sigma1/sigman} and \eqref{sigma2/sigman} into  \eqref{surface impedance}, the surface impedance $Z_s$ can be calculated. 
Then, $\delta_{i,\mathrm{theory}}$ and $n_{qp,\mathrm{theory}}(T)$ can be obtained by:
\begin{equation}
\delta_{i,theory} = \alpha \left(\frac{\text{Real}(Z_s)}{\text{Im}\left(\frac{Z_s}{\omega}\right)}\right) \frac{1}{\omega}
\end{equation}

\begin{equation}
n_{qp,theory}(T) \sim \delta_{qp,theory}(T) \times N_0 \times \Delta(T) \times \frac{\pi}{\alpha} \times \sqrt{\frac{h f_r}{2\Delta(T)}}
\label{n-qp-theory}
\end{equation}
Figure  \ref{Qi-measured-theory}(a) compares the measured and theoretical $Q_i$ values, as indicated by $Q_{i,measured}$ and $Q_{i,theory}$, for temperatures ranging from 77 mK to 1 K. This distinction between $Q_{i,measured}$ and $Q_{i,theory}$ indicates the existence of an additional loss channel originating from the non-equilibrium quasiparticle density. More specifically, we used Eqs. \eqref{n-qp-measured} and \eqref{n-qp-theory} to compute the non-equilibrium quasiparticle density in the CPWs. Theoretically, the quasiparticle density at low temperatures should be negligible, however, measurements indicate that quasiparticle density is present at low temperatures (see the measured and calculated quasiparticle density in Fig. \ref{Qi-measured-theory}(b)). It can be shown in Fig. \ref{Qi-measured-theory}(a) that $Q_{i,measured}$ is lower than predicted, leading us to conclude that at low temperatures, the quasiparticle density stays finite.
In fact, Fig. \ref{Qi-measured-theory}(a) shows $Q_{i,measured}$ and $Q_{i,theory}$ versus temperature at low photon number $\langle n_{ph} \rangle \sim 1$ with the theoretical model of TLS and quasiparticle loss. Both plots are calculated and measured at $f_r = 3.65$ GHz.
Figure \ref{Qi-measured-theory}(c) and (d) demonstrate the relationship between the quasiparticle density 
$n_{\mathrm{qp}}$ and the normalised complex conductivity components, 
$\sigma_{1}/\sigma_{n}$ and $\sigma_{2}/\sigma_{n}$, for the Ta CPW resonator, respectively. Excellent agreement over the measured range is observed when 
the experimental results (red circles) are compared with the Mattis-Bardeen 
theoretical predictions (blue lines). As shown in Fig. \ref{Qi-measured-theory}(c), the value of $\sigma_{1}/\sigma_{n}$ increases in a nearly linear trend as $n_{\mathrm{qp}}$ rises. This behaviour is indicative of the increased dissipative response that results from the increased population of 
unpaired quasiparticles. These quasiparticles contribute to microwave absorption 
through single-particle excitations across the superconducting energy gap. The 
observed increase in the real part of the conductivity is directly attributed to 
the increase in the probability of quasiparticle--photon interactions as 
$n_{\mathrm{qp}}$ is increased.

Conversely, Fig. \ref{Qi-measured-theory}(d) shows a monotonic decline in $\sigma_{2}/\sigma_{n}$ 
as $n_{\mathrm{qp}}$ increases, which is consistent with the anticipated drop in the 
superfluid density. Since the inductive response of the Cooper pair condensate 
determines the imaginary component of the complex conductivity, the generation of 
quasiparticles, whether thermally or through non-equilibrium processes, breaks 
Cooper pairs, reducing the superfluid fraction and, consequently, $\sigma_{2}$. 
In normal superconductors, this negative correlation is a sign of pair-breaking 
dynamics, and it provides a direct means of studying the effects of quasiparticles. 
The quantitative agreement between experiment and Mattis--Bardeen theory over the 
full density range confirms that the electrodynamic response of the resonator is 
well described by the standard microscopic theory without requiring additional 
loss mechanisms.
The experimental results indicate that Ta superconductors exhibit substantially lower quasiparticle densities than NbN superconductors under identical normalised operating conditions. This normalised temperature approach provides a fundamentally more meaningful comparison by evaluating both materials at thermodynamically equivalent operating points, the same fractional distance from their respective superconducting phase transitions. In fact, Ta achieves a remarkably low quasiparticle density of 0.3 $\times 10^{3}$ $(\mu\mathrm{m}^{-3})$  at  $T/T_c$ = 200, which is a threefold decrease from the measured value of $1 \times 10^{3} $ $(\mu\mathrm{m}^{-3})$  for NbN \cite{fischer2023nonequilibrium}. Lower quasiparticle densities directly enhance coherence times and decrease decoherence in qubits by reducing energy dissipation mechanisms. The threefold reduction in quasiparticle density seen in Ta is related to much lower loss tangent values. Moreover, the charge noise and frequency fluctuations that affect superconducting quantum devices, especially single-photon detectors and parametric amplifiers, are significantly reduced by lower quasiparticle populations. For next-generation superconducting quantum technologies, where minimal dissipation is one of the most important topics, Ta is an excellent material choice due to these advantages, as well as its superior material features, including a decreased surface roughness and a reduction in two-level system defects.
\section{Conclusion}
In this work, we provided a comprehensive analysis of high-$Q$ superconducting microwave coplanar waveguide resonators made of $\alpha$-tantalum on silicon with a niobium seed layer. Using transmission electron microscopy and X-ray diffraction, a comprehensive structural investigation was conducted, which showed the production of a mainly $\alpha$-Ta body-centered cubic phase. This phase demonstrated atomically sharp Ta/Nb/Si contacts and minimal interdiffusion, resulting in high film quality for low-loss superconducting applications. In the single-photon regime, temperature-dependent microwave spectroscopy demonstrated the existence of persistent non-equilibrium quasiparticles at millikelvin temperatures. Excess quasiparticles are a leading source of decoherence in superconducting qubits, which significantly reduces qubit coherence times. Longer coherence and enhanced operational fidelity are possible for superconducting devices by inhibiting the formation of quasiparticles and trapping residual excitations. The low intrinsic loss and proven structural quality of $\alpha$-Ta resonators make them ideal building blocks for ultra-sensitive microwave photon sensors, kinetic inductance detectors, and next-generation quantum processors. Our findings demonstrate a consistent electrodynamic behaviour and structural robustness of $\alpha$-Ta that can directly aid in the development of next-generation quantum information systems, where maintaining coherence is the most important performance metric.

\section{Acknowledgements}
This study made use of the University of Glasgow’s James Watt Nanofabrication Centre
(JWNC), and Kelvin Nanocharacterisation Centre (KNC). We thank the technical staff for their
support. This work was supported in part by the Royal Academy of Engineering Fellowship (LTRF2223-19-138), the Royal Society of Edinburgh (319941), Royal Society Research Grant (RGS/R2/222168), EPSRC PRF-11-I-08, and EPSRC EP/X025152/1.

\newpage
\bibliographystyle{ieeetr}

\newpage
\section{Appendix:  
S.1: Microwave spectroscopy setup}
\begin{figure}[H]
    \centering
    \includegraphics[width=0.5\linewidth]{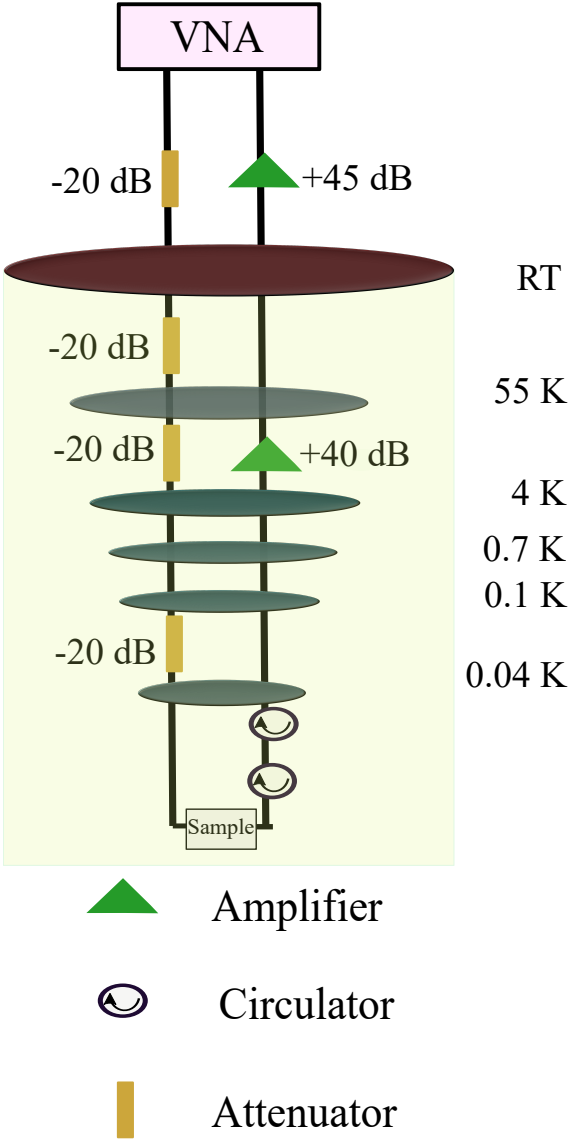}
    \caption{Schematic of the cryogenic setup for sub-Kelvin microwave spectroscopy of the chip.}
    \label{fig:placeholder}
\end{figure}

\end{document}